\documentclass[conference]{IEEEtran}

\usepackage[utf8]{inputenc}
\usepackage[T1]{fontenc}

\PassOptionsToPackage{hyphens}{url}

\usepackage[hyperfootnotes=false, colorlinks=true,
            citecolor=blue, linkcolor=blue, urlcolor=blue]{hyperref}
\usepackage{bookmark}
\usepackage[dvipsnames]{xcolor}

\usepackage{cite}
\usepackage{url}

\usepackage{longtable}
\usepackage{tabularx}
\usepackage[flushleft]{threeparttable}
\usepackage{float}
\usepackage{multirow,booktabs}
\usepackage{makecell}
\usepackage{graphicx}
\usepackage{adjustbox}
\graphicspath{ {./images/} {./} }
\usepackage{caption}
\usepackage{subcaption}
\usepackage{comment}

\usepackage{amsmath}
\usepackage{array}
\usepackage{amsthm}
\usepackage{amssymb}

\theoremstyle{definition}

\usepackage{tikz}
\usetikzlibrary{shapes, arrows, positioning, calc}

\usepackage{hyphenat}
\usepackage[bottom]{footmisc}
\usepackage{csquotes}
\usepackage{pdfpages}
\usepackage{pdflscape}

\begin{document}

\title{Burnyard: Future of Malware Analysis}

\author{
\IEEEauthorblockN{Rama Ramana Sharma Parnandi}
\IEEEauthorblockA{
    \small{parnandi.2@osu.edu}\\
    \small{College of Engineering}\\
    \small{The Ohio State University}\\
    Columbus, OH, USA
}
\and
\IEEEauthorblockN{Carter Yagemann}
\IEEEauthorblockA{
    \small{yagemann.1@osu.edu}\\
    \small{College of Engineering}\\
    \small{The Ohio State University}\\
    Columbus, OH, USA
}
}

\maketitle

\begin{abstract}
Malware analysis is a critical aspect of modern cybersecurity. The prevailing industry practice, sandboxing, involves executing suspicious binaries within isolated virtual machines in large-scale data centers. However, this approach can unintentionally expose samples to public platforms such as VirusTotal and MalwareBazaar, and it is both resource-intensive and time-consuming.

Burnyard addresses these limitations through a lightweight binary emulation platform that captures observable runtime behavior and records it as structured CSV event traces.
\end{abstract}


\section{INTRODUCTION}
The evolution of malware over the past decades demonstrates a significant increase in both sophistication and impact. In 2000, the ILOVEYOU virus, created by a 24-year-old resident of the Philippines, spread globally with the aim of harvesting user credentials. Its author, Onel de Guzman, reportedly developed the malware to gain unauthorized internet access ~\cite{ILOVEYOU}. The attack caused an estimated \$3 to \$15 billion in damages worldwide ~\cite{ebsco}.

Nearly a decade later, in 2013, CryptoLocker ransomware emerged, notable for its rapid spread and use of asymmetric encryption. It was distributed through botnets linked to the Zeus Trojan, which served as a delivery mechanism for multiple types of malware. By 2016, the Mirai botnet marked another shift in attack strategies by targeting Internet of Things (IoT) devices, creating a large-scale botnet responsible for one of the most significant distributed denial-of-service (DDoS) attacks on record ~\cite{ILOVEYOU}.

By 2026, malware has become significantly more advanced, with certain families employing techniques to evade detection by modifying or concealing their behavior during analysis. The widespread adoption of artificial intelligence following the release of ChatGPT in late 2022 has further reshaped the threat landscape, enabling adversaries to automate and refine malware development. In November 2025, researchers at NYU Tandon introduced PromptLock, an AI-driven proof-of-concept ransomware~\cite{nyutandon}, raising important concerns about the extent to which malicious actors can exploit AI to carry out sophisticated attack chains.

Malware analysis techniques are broadly classified into static and dynamic approaches. Static analysis involves examining a binary without executing it, with a focus on its code structure, embedded strings, system calls, and metadata. In contrast, dynamic analysis entails running the malware in a controlled environment to observe its runtime behavior, including system interactions and network activity.

At an industrial scale, dynamic analysis is predominantly implemented through sandboxing~\cite{Fortinet}. This approach relies on large data center infrastructures to host diverse virtual environments for analyzing suspicious binaries. However, sandboxing is both resource-intensive and time-consuming. Additionally, uploading samples to public platforms such as VirusTotal or MalwareBazaar can unintentionally expose them to broader access.

Modern malware increasingly employs advanced evasion techniques, such as polymorphism, obfuscation, and anti-analysis mechanisms, which weaken the effectiveness of traditional signature-based detection methods. Although sandboxing provides valuable behavioral insights, it remains vulnerable to evasion by malware capable of detecting virtualized environments, often preventing analysts from observing its true intent.

\section{BACKGROUND}\label{Background}

This section provides a concise overview of the key theoretical concepts and prior research that underpin this project.

\subsection{Malware}

Malware is broadly defined as any software engineered to disrupt systems, compromise data, gain unauthorized access, or cause harm to individuals or organizations. Based on shared behavioral traits and code lineage, malware is commonly grouped into families. Major categories include remote access trojans (RATs), which enable persistent remote control; droppers and loaders, which deliver secondary payloads; botnets, which orchestrate compromised devices for large-scale operations; ransomware, which encrypts data to extort payment; and trojans, which disguise themselves as legitimate software while carrying out malicious actions.

\subsection{Static Analysis}

Static analysis involves examining a binary without executing it. Analysts assess elements such as code structure, embedded strings, import tables, section entropy, and metadata to identify indicators of malicious behavior, obfuscation, or known signatures. While static analysis is efficient and does not require an execution environment, its effectiveness is limited when dealing with packed or heavily obfuscated samples, where the underlying code is only revealed at runtime.

\subsection{Dynamic Analysis}

Dynamic analysis involves executing a program within a controlled environment to observe its runtime behavior. This approach captures system-level interactions such as file system activity, registry modifications, network communications, process creation, and memory usage that are not accessible through static analysis alone.

Dynamic techniques are particularly effective against obfuscated or packed malware, as such samples must eventually unpack and execute their true payload, thereby revealing their malicious behavior. However, dynamic analysis requires an execution environment that is both sufficiently realistic to elicit authentic behavior and adequately isolated to prevent harm to production systems.

\subsection{Sandboxing}

Sandboxing is the dominant industrial approach to dynamic malware analysis. It provides an isolated execution environment; typically a full virtual machine in which a sample is executed and its behavior is monitored. To prevent any lasting impact on the host system, the environment is reset or discarded after analysis. At scale, platforms such as Cuckoo~\cite{cuckoosandbox} and cloud-based services like VirusTotal rely on large pools of virtual machines to process submitted samples. While effective, this approach is resource-intensive, demands substantial infrastructure, and may expose samples to third-party systems, making it less suitable for air-gapped or privacy-sensitive environments.

\subsection{Syscalls and API Calls as Behavioral Features}

A system call (syscall) is a mechanism through which a user-space application requests the operating system kernel to perform privileged operations, such as file access, network communication, or process management. In Windows environments, applications generally interact with the operating system via the Win32 API, which ultimately maps to lower-level native system calls. The sequence of system and API calls generated during execution forms a program’s behavioral fingerprint. Distinct malware families exhibit characteristic patterns: ransomware typically involves intensive file and cryptographic operations; remote access trojans (RATs) emphasize network communication and process injection; and droppers focus on payload delivery and execution activities.

\section{RELATED WORK}\label{Related Work}

This section reviews prior research related to the three foundational components of Burnyard: binary emulation as an analysis framework, system call and API trace analysis for malware classification, and sandbox evasion as a significant challenge to dynamic analysis systems.

\subsection{Binary Emulation for Malware Analysis}

The work of Vouvoutsis et al.~\cite{borzacchiello2022emulation} closely aligns with the objectives of Burnyard by investigating the viability of binary emulation as an alternative to traditional sandbox-based malware analysis for classification tasks. In their study, a dataset of 71,536 binaries was emulated, with API call sequences extracted as behavioral features and used to train machine learning models for malware family classification. The findings indicate that emulation-derived traces can achieve classification performance comparable to that of conventional sandboxing approaches, while eliminating the need for a full virtual machine environment.

However, their methodology relies exclusively on built-in API tracing without incorporating custom hook implementations. As a result, the feature set is constrained to the APIs supported at the time of the study. The authors acknowledge this limitation, noting that incomplete coverage of APIs and system calls can prevent certain binaries from completing full emulation. This, in turn, restricts the depth and completeness of the collected data, which they characterize as only a partial representation of true runtime behavior~\cite{borzacchiello2022emulation}.

This limitation is reflected in the feature space: although 2,536 features were initially extracted, only 239 remained after feature selection, indicating substantial sparsity due to incomplete API coverage.

\subsection{Syscall and API Call Based Classification}
A recent survey by Gond and Mohapatra~\cite{carrier2025syscall} provides a comprehensive review of malware detection techniques based on system and API call analysis across Linux, Windows, and Android platforms. Leveraging Cuckoo Sandbox to capture API activity, the study evaluates static, dynamic, and hybrid approaches, identifying API call sequences as highly discriminative features for malware classification.

The SaMOSA system~\cite{samosa2025} adopts a related methodology by analyzing syscall frequency distributions alongside network activity observed during sandbox execution. By incorporating multiple side-channel signals, the approach improves detection robustness and enhances classification accuracy.

Earlier work by Ahmed et al.~\cite{ahmed2022joint} further demonstrates the effectiveness of combining static binary features with dynamic API traces. Their joint analysis framework utilizes transfer learning to develop efficient models that generalize well across diverse malware families.

\subsection{Sandbox Evasion and Anti-Analysis Countermeasures}

A persistent challenge in dynamic malware analysis is that advanced samples actively probe their execution environment and alter or suppress their behavior when analysis conditions are detected. Ruggia et al.~\cite{ruggia2024evasive} report that approximately 70\% of evasive Android malware targets emulated environments, employing techniques such as timing analysis, CPU feature inspection, and user-interaction checks.

Similar evasion strategies are also observed in Windows malware, highlighting a well-documented limitation of sandbox-based analysis. In such environments, the presence of instrumentation or virtualization artifacts can inhibit the observation of genuine malicious behavior, thereby reducing the effectiveness of dynamic analysis approaches~\cite{miramirkhani2017sandboxevasion}.

\subsection{Machine Learning for Malware Detection}

Beyond API call sequences, researchers have explored a wide range of feature representations for malware classification. Gaber et al.~\cite{gaber2021malwaresurvey} present a comprehensive survey of AI-driven malware detection techniques, including approaches based on raw byte analysis, control flow graphs, n-gram features, and learned neural embeddings. Collectively, these methods illustrate the diversity of feature engineering strategies available for improving classification accuracy and robustness.

\subsection{Positioning of Burnyard}

Established tools such as Cuckoo Sandbox~\cite{cuckoosandbox} and commercial platforms like VirusTotal represent the current industry standard for dynamic malware analysis. These systems rely on full virtual machine execution, requiring substantial computational infrastructure, and may expose submitted samples to third-party access or public visibility.

Burnyard adopts a different design approach, trading the complete behavioral fidelity of a full operating system environment for the operational benefits of user space emulation; namely reduced infrastructure requirements, improved architectural portability, and enhanced sample privacy.

\section{SYSTEM ARCHITECTURE} \label{System Architecture}

\begin{figure}[!ht]
\centering
\begin{tikzpicture}[
    node distance=0.5cm,
    box/.style={rectangle, rounded corners=3pt, minimum width=3.4cm, minimum height=0.75cm,
                text centered, align=center, font=\small},
    arr/.style={->, >=stealth, semithick, gray!70},
]

\node[box, fill=blue!10,   draw=blue!50]  (upload)  {Submit Binary\\[-1pt]{\tiny PE / ELF / JS / BAT / PS1 \ldots}};
\node[box, fill=violet!10, draw=violet!50, below=of upload]  (static)  {Metadata Collection};
\node[box, fill=green!10,  draw=green!60, below=of static]   (tracer)  {Tracer};
\node[box, fill=orange!10, draw=orange!50,below=of tracer]   (csv)     {Event Trace};
\node[box, fill=teal!8,    draw=teal!40,  below=of csv]      (pred)    {Prediction Pipeline};
\node[box, fill=red!10,    draw=red!40,   below=of pred]     (report)  {Analysis Report};

\draw[arr] (upload) -- (static);
\draw[arr] (static) -- (tracer);
\draw[arr] (tracer) -- (csv);
\draw[arr] (csv)    -- (pred);
\draw[arr] (pred)   -- (report);

\end{tikzpicture}
\caption{Burnyard end-to-end analysis workflow.}
\label{fig:workflow}
\end{figure}
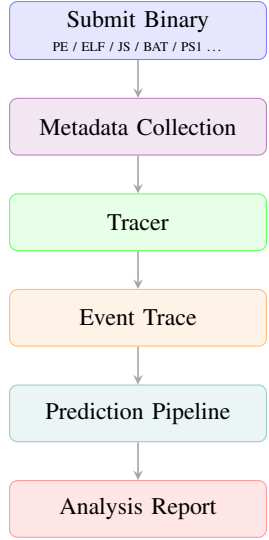

This section presents the technical design and implementation of Burnyard. The system is composed of four tightly integrated components: a binary tracer, a syscall and API hook framework, a machine learning classification pipeline, and a web application that orchestrates the full analysis workflow. The overall analysis workflow of Burnyard is shown in Fig.~\ref{fig:workflow}.

\subsection{Overview}

Burnyard's workflow is divided into two phases. In the first phase, the submitted binary is executed inside an emulated environment paired with a corresponding root filesystem. During emulation, every syscall and Windows API call issued by the program is intercepted by a custom hook handler and persisted as an event trace. In the second phase, the classification pipeline consumes that trace and assigns the sample a label which is either benign or a known malware family.

Burnyard's emulation layer operates at the user-space instruction level rather than requiring a full hypervisor stack, enabling deployment on resource-constrained hardware without network connectivity. Also, the submitted samples are never transferred to any external service, ensuring complete analysis privacy.

\subsection{Tracer}

The tracer module is the entry point for dynamic analysis. It accepts a binary and a root filesystem directory, and produces a structured record of all observable runtime activity syscalls, API calls, parameters, and return values that serves as input to the classification pipeline.

At startup, the tracer configures the emulation engine with the operating system and CPU architecture profile appropriate for the target binary. Burnyard supports cross-platform emulation across Windows, Linux, and Mach-O targets, spanning multiple CPU architectures. The filesystem supplies the libraries, system directories, and registry stubs that the binary expects at runtime, eliminating the need for a genuine host operating system.

\subsection{Syscall and API Hook Framework} \label{SyscallHooks}

Every syscall and Windows API call issued during emulation is intercepted by the hook framework, which writes as an structured event. The tracer emits one output row per intercepted event. Each row contains the event name (e.g., \texttt{CreateFileA}, \texttt{read}), decoded parameters (resolved file paths, flag values, error codes), and the return value. Raw memory values and integer constants are translated into human-readable strings by per-hook decoding helpers. The resulting trace is a flat, chronologically ordered record of all observable behavior and serves as the direct input to the classification pipeline.

\subsection{Classifier}
Based on the events captured and data points collected by the tracer the given sample is classified either into benign or a known malware family. A transformer-based language model (SLM) also produces a natural-language behavioral explanation.

\section{RESULTS}\label{Results}

This section evaluates Burnyard’s analysis throughput in comparison with two established platforms: VirusTotal and Sophos Intelix. The primary metric is the average analysis time per sample, measured across 100 samples for each operating system category.

VirusTotal is a multi-vendor aggregation platform that submits each sample to more than 70 security engines in parallel and returns a consolidated verdict. Most of these engines rely on static analysis examining file structure, embedded strings, import tables, and known signatures rather than executing the sample in a controlled environment. While a subset of engines performs dynamic analysis via sandboxing, the reported result reflects the aggregate response time across all engines. As a result, VirusTotal’s latency is primarily driven by static scanning performance rather than sandbox execution time.

Intelix, a threat intelligence API developed by Sophos, adopts a different approach. Unlike VirusTotal, Intelix provisions a dedicated sandbox environment for each submitted sample, executing it to observe runtime behavior before issuing a verdict. While this per-sample sandboxing provides thorough behavioral analysis, it introduces overhead associated with initializing, executing, and tearing down an isolated environment for each submission.

\subsection{Analysis Time Comparison}

Table~\ref{tab:timing} presents the average end-to-end analysis time, measured in seconds, for Windows PE and Linux ELF samples across each platform, based on a combined set of 100 samples. For Burnyard, this metric includes the full analysis pipeline, encompassing metadata extraction, emulation, trace generation, and classification.

\begin{table}[h]
\centering
\caption{Average analysis time (seconds) per platform across 100 combined samples}
\label{tab:timing}
\begin{tabularx}{\linewidth}{>{\raggedright\arraybackslash}X
                              >{\centering\arraybackslash}X
                              >{\centering\arraybackslash}X
                              >{\centering\arraybackslash}X}
\hline
\textbf{OS Type} & \textbf{Intelix} & \textbf{VirusTotal} & \textbf{Burnyard} \\
\hline
Windows & 182.88 & 32.36 & 22.41 \\
Linux   & 80.85  & 16.27 & 5.47  \\
\hline
\end{tabularx}
\end{table}

\subsection{Discussion}
For the purposes of this evaluation, Burnyard was deployed on a Dell Optiplex Micro 3050 equipped with a 7th-generation Intel i5 processor and 16~GB of DDR3 RAM. Burnyard achieves the lowest average analysis time across both operating system categories. For Windows samples, Burnyard completes analysis in an average of 22.41 seconds, compared to 32.36 seconds for VirusTotal and 182.88 seconds for Intelix representing a $1.44\times$ speedup over VirusTotal and an $8.16\times$ improvement over Intelix.

The performance advantage is even more pronounced for Linux samples, where Burnyard averages 5.47 seconds, compared to 16.27 seconds for VirusTotal and 80.85 seconds for Intelix. This corresponds to speedups of $2.97\times$ and $14.78\times$, respectively.

The observed latency gap between VirusTotal and Intelix reflects the fundamental differences in their analysis approaches. VirusTotal's relatively low response time is primarily driven by its reliance on static analysis across most contributing engines, which does not require sample execution. In contrast, Intelix provisions a dedicated sandbox environment for each submission, and the overhead associated with environment initialization, execution, and teardown is reflected in its significantly higher average analysis time; 182.88 seconds for Windows and 80.85 seconds for Linux.

Burnyard's greater performance margin over Intelix for Linux samples, compared to Windows samples, is consistent with the lighter-weight nature of ELF binaries. Linux samples typically produce shorter execution traces, enabling faster processing within the classification pipeline. Conversely, Windows samples involve a broader Win32 API surface and more complex dynamic linking (DLL resolution), which moderately increases both emulation and classification time.

Importantly, Burnyard achieves these results without reliance on external network connectivity or cloud infrastructure. Unlike VirusTotal and Intelix, which process submitted samples through remote pipelines and retain artifacts on third-party systems, Burnyard operates entirely on the local host. This eliminates network-induced latency and ensures that samples remain confined to the analyst's environment; a property not provided by either VirusTotal or Intelix.

\subsection{Classification Accuracy}

\begin{figure}[!ht]
\centering
\includegraphics[width=\linewidth]{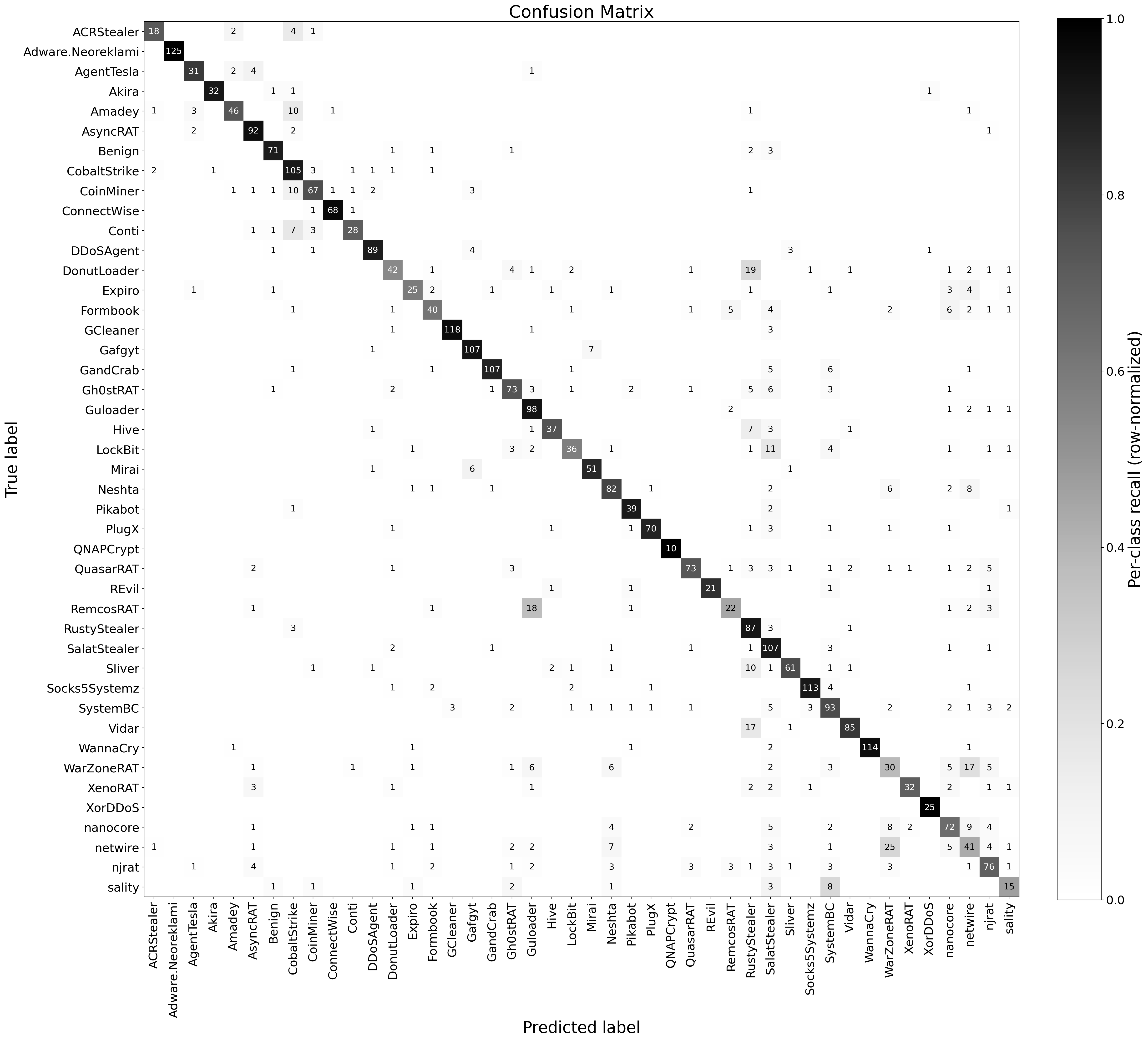}
\caption{Confusion matrix across 44 classes (row-normalized). Diagonal shading indicates per-class recall.}
\label{fig:confusion}
\end{figure}

Fig.~\ref{fig:confusion} illustrates the confusion matrix generated by the classification pipeline, evaluated across 44 classes comprising 43 distinct malware families and one benign class. The matrix is row-normalized, with cell shading representing per-class recall, and absolute prediction counts overlaid for reference.

The matrix's significant diagonal dominance shows that most classes have high recall. Well-represented families with greater sample sizes, such as Adware.Neoreklami (125), GCleaner (118), WannaCry (114), Socks5Systemz (113), and CobaltStrike (105), have consistently good recall, as indicated by the darker diagonal entries.

In contrast, several families exhibit lower recall, including QNAPCrypt (10), salty (15), REvil (21), and RemcosRAT (22). These cases correspond to classes with limited training samples, where insufficient behavioral diversity in the training data constrains the model's ability to generalize effectively.

Notably, the remaining misclassifications are primarily concentrated among semantically or behaviorally related families, rather than being randomly distributed across the label space. This pattern is consistent with the underlying behavioral feature representation, where similar operational characteristics can lead to classification ambiguity.

\textit{Ransomware:} LockBit and Hive exhibit mutual confusion, as both generate encryption-intensive cryptographic API sequences with similar file enumeration and deletion behaviors. Conti and Akira demonstrate minor cross-confusion for similar reasons. In contrast, WannaCry achieves near-perfect recall, attributable to its distinctive propagation mechanism leveraging SMB exploitation, which differentiates its behavioral profile.

\textit{Remote Access Trojans (RATs):} The RAT cluster comprising WarZoneRAT, njrat, nanocore, netwire, QuasarRAT, RemcosRAT, and XenoRAT accounts for a significant portion of off-diagonal mass. These families share overlapping behavioral characteristics, including persistent process injection, keylogging-related API usage, and outbound command-and-control (C2) communication. WarZoneRAT is most frequently misclassified as njrat, while nanocore exhibits partial overlap with netwire; both pairs share similarities as .NET-based RATs with comparable runtime call patterns.

\textit{Stealers:} Stealer families, including ACRStealer, RustyStealer, SalatStealer, and Vidar, are generally well-separated, with misclassifications limited to a small number of samples. Their behavioral traces are primarily distinguished by variations in browser credential access and cryptocurrency wallet enumeration API sequences, which provide sufficient differentiation for reliable classification.

\textit{Botnets and DDoS:} Botnet and DDoS-oriented families; such as Mirai, Gafgyt, DDoSAgent, Socks5Systemz, and XorDDoS form a distinct cluster. Mirai and Gafgyt, both Linux ELF-based IoT botnets, exhibit slight mutual confusion due to similar socket-intensive network behavior. However, differences in their propagation-related syscall patterns enable the classifier to distinguish between them in most cases.

Overall, the confusion matrix indicates that the classification pipeline achieves robust family-level identification across a large and taxonomically diverse malware corpus. Misclassifications are not random but structurally aligned, occurring primarily among families that share common behavioral primitives. This pattern demonstrates that the extracted event trace features capture meaningful intra-class characteristics while preserving discriminative signals across malware families.

\section{CONCLUSION}\label{Conclusion}

This paper presents Burnyard, a self-contained malware analysis system that integrates user-space binary emulation with a comprehensive system call and API hooking framework to generate detailed behavioral event traces. These traces are subsequently processed by a custom classification pipeline for malware identification. The system supports Windows, Linux, and Mach-O binaries across multiple CPU architectures and operates without requiring network connectivity, virtual machine infrastructure, or reliance on external services. For accelerated prediction, a low-end GPU can optionally be utilized.

Evaluation against two established platforms demonstrates that Burnyard achieves lower average analysis times across both Windows and Linux datasets. For Windows binaries, Burnyard is $1.44\times$ faster than VirusTotal and $8.16\times$ faster than Intelix. For Linux binaries, the performance advantage increases to $2.97\times$ over VirusTotal and $14.78\times$ over Intelix. These improvements are achieved without compromising analysis privacy, as all samples remain on the analyst's local system throughout the process, addressing a key limitation of cloud-based sandboxing solutions.

Beyond performance, Burnyard's architecture provides a significant operational advantage for air-gapped and resource-constrained environments. By operating at the user-space emulation layer rather than relying on hypervisor-based virtual machines, the system can be deployed on commodity hardware while maintaining independence from external infrastructure.

\bibliographystyle{IEEEtran}
\bibliography{biblography.bib}

\end{document}